\documentclass[superscriptaddress,twocolumn,aps,prl,floatsfix]{revtex4}
\usepackage{graphics}
\usepackage{amssymb,amsmath}
\usepackage{array}
\usepackage{dcolumn}

\begin{document}

\title{The role of dynamical polarization of the ligand to metal
charge transfer excitations in {\em ab initio} determination 
of effective exchange parameters} 

\author{Alain Gell\'e}
\affiliation{Laboratoire de Physique Quantique, IRSAMC~/~UMR~5626,
Universit\'e Paul Sabatier, 118 route de Narbonne, F-31062 Toulouse
Cedex 4, FRANCE}

\author{Marketa L. Munzarov\'a} 
\affiliation{National Center for Biomolecular Research, Masaryk University, 
Faculty of Sciences, Kotlarska 2, CZ- 611 37 Brno, Czech Republic}

\author{Marie-Bernadette Lepetit} 
\affiliation{Laboratoire de Physique Quantique, IRSAMC~/~UMR~5626,
Universit\'e Paul Sabatier, 118 route de Narbonne, F-31062 Toulouse
Cedex 4, FRANCE}

\author{Francesc Illas}
\affiliation{Departament de Qu\'{\i}mica F\'{\i}sica \&
Centre especial de Recerca en Qu\'{\i}mica Te\'orica, Universitat de
Barcelona \& Parc Cientific de Barcelona, C/ Mart\'{\i} i Franqu\'es 1,
08028 Barcelona, SPAIN}   

\date{\today}

\begin{abstract}

The role of the bridging ligand on the effective Heisenberg coupling
parameters is analyzed in detail. This analysis strongly suggests that
the ligand-to-metal charge transfer excitations are  responsible
for a large part of the final value of the magnetic coupling
constant. This permits to suggest a new variant of the Difference
Dedicated Configuration Interaction (DDCI) method, presently one of
the most accurate and reliable for the evaluation of magnetic
effective interactions. This new method treats the bridging ligand
orbitals mediating the interaction at the same level than the magnetic
orbitals and  preserves the high quality of the DDCI results while
being much less computationally demanding.
The numerical accuracy of the new approach is
illustrated on various systems with one or two magnetic electrons per
magnetic center. The fact that accurate results can be obtained using
a rather reduced configuration interaction space opens the possibility
to study more complex systems with many magnetic centers and/or many
electrons per center.

\pacs{71.15.-m, 74.25.Ha, 74.72.-h, 75.10.Jm }
\end{abstract}

\maketitle

\section{Introduction}
\label{sec:Intro}

Strongly correlated systems have attracted a lot of attention in the
last decade. Indeed, chemists have synthesized a large number of materials 
and new families of molecular systems which present unusual and
fascinating properties directly related to the strongly correlated
character of their electronic structure. One can cite as examples the
high $T_c$ superconducting copper oxides~\cite{htc}, colossal
magneto-resistant manganite oxides~\cite{mang}, photo-magnetic
polycyanides molecules~\cite{phmgmol} and materials~\cite{phmg}, molecular 
magnets~\cite{book-olivier-kahn}, etc.

In all these materials a few electrons (per unit cell) are responsible
for their spectacular properties. These electrons are usually unpaired
and localized both spatially and energetically near the Fermi level
(from now on we will refer to them as {\em magnetically active
electrons}).  Consequently, the electronic wave function of these
systems is essentially multi-configurational and cannot be correctly
treated by single-reference based methods such as Hartree-Fock plus
perturbation theory or even Density Functional Theory~\cite{dft}. The
description and rationalization of the physical properties of these
systems is usually carried out in terms of effective valence-bond (VB)
type Hamiltonians describing the interactions between the above
defined {\em magnetically active electrons}~\cite{revue,HEFF}.
Among the most widely used effective models we quote the
Heisenberg~\cite{heis} model ---~which describe the effective exchange
between unpaired electrons~--- or the $t-J$~\cite{t-J} model ---~which
treats in addition to the exchange phenomenon the hole or electron
hopping. Clearly, properties such as the metallic versus insulating
character of a given compound, the total magnetization or the magnetic
order, depend on the relative amplitudes of the different effective
integrals. Therefore, it is of crucial importance to be able
to quantify the different effective terms of these
models. Unfortunately, experimental data are often unable to determine
these effective parameters and, for this purpose, it is necessary to rely on 
accurate {\em ab-initio} quantum-chemical calculations~\cite{CALZADO99}.

Here, it is worth to mention that in recent years a large amount
of quantum chemical calculations are nowadays carried out within the
framework of density functional theory (DFT). However, it is well
known that these methods badly fail in the treatment of such
strongly correlated systems.  For instance, DFT fails even to predict
the insulating character of NiO~\cite{NiO}. Also, one needs to recall
that DFT methods tend to largely overestimate the singlet triplet
local excitation energy associated with the magnetic coupling in the 
high $T_c$ cuprate super-conductors~\cite{dft1}. For the
$L\!a_2C\!uO_4$ compound, the Local Density Approximation (LDA)
estimates of the local magnetic exchange range from $600meV$ to
$800meV$ according to the functional used. The experimental
measurements~\cite{la2cuo4} yield $135\pm 6meV$. For such problems it
is therefore necessary to rely on wave-function, multi-reference,
ab-initio, spectroscopy methods such as the Difference Dedicated
Configurations Interaction~\cite{ddci} (DDCI) which has proved to be
able to quantitatively reproduce the magnetic coupling constant of a
large family of magnetic systems~\cite{CALZADO99}. The DDCI method
uses a configurations interaction expansion of the electronic wave
function which is specially designed for the calculation of excitation
energies involving essentially open-shell states. This configuration
interaction (CI) method is based on the choice of the so-called
valence complete active space (CAS), on which a CI expansion is
constructed including all possible Slater determinants belonging to the
CAS, plus a selection of the single and double excitations out the CAS
determinants. This selection is done in order to eliminate from the
Hamiltonian matrix all configurations that do not contribute to the
considered energy difference at the second order of perturbation
theory, that is all double excitations from two inactive orbitals
toward two virtual ones.

The DDCI based approaches have been proved to yield very accurate
results for the calculation of excitation energies in magnetic
systems, both for molecular spectroscopy and local excitations in
solids. On top of the results already mentioned~\cite{CALZADO99}, 
one can cite the remarkable accuracy of the
calculation of the singlet triplet excitation energy in
oxalato-bridged $C\!u(II)$ binuclear complexes, where the error of the
DDCI method is smaller than $5{\rm cm}^{-1}$ compared to the
experimental data~\cite{molmagn}, 
the prediction of the magnetic exchange in super-conductor copper
oxides~\cite{xino} or the on-rung doublet-doublet excitation on the
famous $\alpha^\prime N\!aV_2O_5$~\cite{vana1}.

While the DDCI method offers the possibility of accurate determination of 
effective parameters for strongly correlated systems it has a rather
serious drawback. This is the dimension of the Hamiltonian matrix
to be diagonalized with a concomitant huge numerical cost. This problem
may even render the calculation unfeasible for systems with too many open
shells arising from several magnetic atoms, several open-shell orbitals per
center or too large ligands. Indeed, the dimension of the CI space scales
as $n_{CAS} \times n_{norb}^2 \times n_{occ}$ where $n_{CAS}$ is the number of 
determinants in the complete active space, CAS, used as reference space,
$n_{norb}$ and $n_{occ}$ refer 
respectively to the total number
of orbitals and the number of inactive orbitals. It would therefore be
highly desirable to determine the key contributions to the energy
difference within the DDCI space and hence be able to restrict the
diagonalization space to these configurations without significant loss of
accuracy. It is clear that such
an analysis may be system-type and excitation-type dependent. Nevertheless, 
a detailed analysis of the different physical contributions strongly suggests
that some general rules exists. In this
paper we will describe these rules by examining the magnetic coupling 
parameters of various bridged bimetallic systems, either molecular or embedded 
cluster models of magnetic materials with localized magnetic moments.

The next section analyzes the physics of the through-bridge
effective magnetic exchange related to the singlet-triplet 
excitation. In view of this analysis, section 3 describes
the physical content of the DDCI wave function, as well as some simplified
versions such as DDCI2 and CAS+single and propose an alternative method to
DDCI. Section 4 proposes a numerical criterion for the
selection of the pertinent ligand orbitals to be included in the
alternative method and will present a systematic numerical study of
both methods on a series of copper- and nickel-based
compounds. Finally section 5 reports the conclusions and
perspectives.

\section{On the bridging ligand role} 
\label{sec2}

The locality of the effective magnetic exchange integrals has been
largely discussed in the literature~\cite{revue,xinoloc,xinotca}. It
has been shown, both theoretically and numerically, that it can be
described within a system fragment involving only i) the two centers
supporting the {\em magnetically active electrons} (from now on
denoted as the {\em magnetic centers}), ii) the bridging ligands and
iii) the first shell of neighboring closed-shell chemical entities
responsible for the screening of the interaction. It is clear that for
an analysis of the interactions within the valence shell, only the
{\em magnetic centers}, associated with their {\em magnetic orbitals}
supporting the {\em magnetically active electrons}, and the bridging
ligands need to be considered.

\begin{figure}[h]
\resizebox{9cm}{4.3cm}{\includegraphics{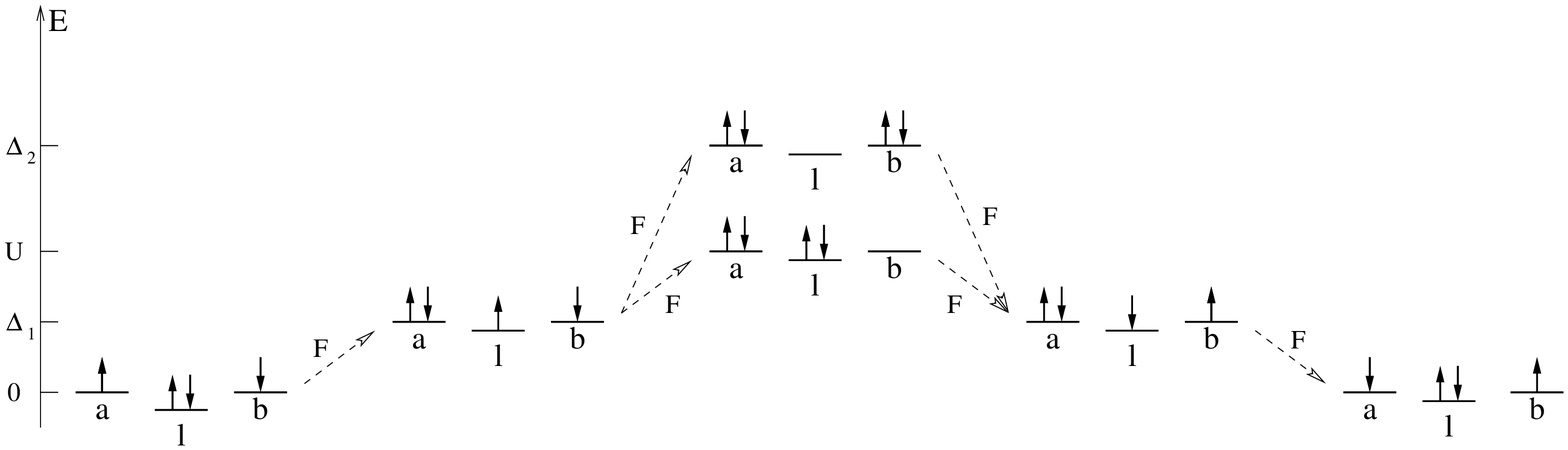}}
\vspace*{0.5eM}
\caption{Through-bridge effective exchange mechanism.}
\label{f:4or}
\end{figure}

Let us first suppose that each {\em magnetic center} supports one
unpaired electron. The effective exchange ${\cal J}_{ab}$ between
these {\em magnetically active electrons} can be analyzed using
quasi-degenerated perturbation theory~\cite{qdpt,HEFF}. The {\em
magnetically active electrons} being strongly correlated the most
natural model space is composed of the two neutral determinants
$|\left(\prod_i \varphi_i^2\right) a \bar b\rangle$ and
$|\left(\prod_i \varphi_i^2\right) b \bar a\rangle$, where $a$ and $b$
are the two {\em magnetic orbitals} respectively associated with the
two {\em magnetic centers} $A$ and $B$, and $\varphi_i$ stands for any
of the doubly-occupied orbitals.  Clearly, the strongly correlated
character of this problem necessitates that correlation effects be
accounted for in the zeroth-order Hamiltonian, while the
delocalization can be described in the perturbative part. For this
reason we will assume an Epstein-Nesbet~\cite{en} partition of the
Hamiltonian and a zeroth-order wave function written in terms of the
localized above cited determinants.  In addition, the two {\em
magnetic centers}, being bridged by closed-shell ligands, are
separated by a rather large distance. Therefore, the direct
interaction between them, such as the direct exchange integral
$K_{ab}$ and the through-space hopping integral $t_{ab}$ can be
considered as negligible in front of the through-bridge
interactions. As a consequence the through-space Anderson's
super-exchange mechanism, that scales as $4t_{ab}^2/U$ is negligeable
in such bridged systems. 
We will consequently consider only the through-bridge interactions in the
forthcoming analysis.  The leading term of the effective exchange
integral thus results from the fourth perturbative order (see
fig.~\ref{f:4or}) and can be expressed as a function of a
bridging-ligand orbital $l$,
\begin{eqnarray} \label{eq:J}
{\cal J}_{ab} &=& -2 {\left(F_{al}^{(al)}\right)^2  \left(F_{al}^{(l)}\right)^2
  \over \left(\Delta_1\right)^2  U} 
-4{\left(F_{al}^{(al)}\right)^2 \left(F_{al}^{(b)}\right)^2 \over 
\left(\Delta_1\right)^2  \Delta_2}
\end{eqnarray}
where $F$ stands for the appropriate Fock operator and $H$ for the Hamiltonian. If
$C$ symbolizes the remaining doubly-occupied orbitals $\left(\prod_{i\ne
l} \varphi_i^2 \right)$ and $a$ and $b$ are supposed symmetry related, one has 
\begin{eqnarray*}
F_{al}^{(al)} &=& \langle C\, a l \bar l \bar b|F| C\, a l \bar a  \bar b \rangle \\
F_{al}^{(l)} &=& \langle C\, b \bar b l \bar l |F| C\,  b \bar b l \bar a \rangle \\
F_{al}^{(b)} &=& \langle C\, b \bar b a \bar l |F| C\, b \bar b a \bar a \rangle \\
\Delta_1 &=& \langle C\, a \bar a l \bar b |H|  C\, a \bar a l \bar b \rangle - 
\langle C\, a  l\bar l \bar b |H|  C\, a l  \bar l \bar b \rangle \\
U &=& \langle C\, a \bar a l \bar l |H|  C\, a \bar a l \bar l \rangle - 
\langle C\, a  l\bar l \bar b |H|  C\, a l  \bar l \bar b \rangle \\
\Delta_2 &=&  \langle C\, a \bar a b \bar b |H|  C\, a \bar a b \bar b \rangle  - 
\langle C\, a  l\bar l \bar b |H|  C\, a l  \bar l \bar b \rangle
\end{eqnarray*}

In the case of $S=1$ atoms, such as $N\!i^{2+}$, the two
(quasi-)degenerated $e_g$-like orbitals support two
ferromagnetically-coupled unpaired electrons. The exchange integral
involves then the $|S_z(A) = 1~; S_z(B) = O\rangle$ and $|S_z(A) = 0~;
S_z(B) = 1\rangle$ configurations. In this case two ligand orbitals
have to be involved, respectively coupling the two types of $e_g$
orbitals (let us say $e_g(1)$ and $e_g(2)$).  It can be simply shown
that i) the effective exchange integral is still equal to the singlet
triplet excitation energy, ii) the above mechanism is still valid for
each of the $e_g$ orbitals and iii) the total exchange can be
expressed as ${\cal J}_{ab} = 1/2 \; {\cal J}_{ab}(e_g(1)) + 1/2\;
{\cal J}_{ab}(e_g(2))$.

From the above expressions it results that the first order correction
to the singlet $|\psi_0(S\!g)\rangle = \left(|C\,l^2a\bar b\rangle +
|C\,l^2b\bar a\rangle \right)/\sqrt{2}$ state is due to the ligand to
metal charge transfer configurations, the second order correction
resulting from the double ligand-to-metal charge-transfer and the
metal-to-metal charge transfer configurations. These first and second
order corrections are highly related to the perturbative evaluation of
the above effective exchange since
\begin{align*}
|\psi_1&(S\!g)\rangle = \\&-\sqrt{2} \; {F_{al}^{(al)} \over \Delta_1 } \; 
{|C\,a^2 l \bar b\rangle + |C\,a^2 b \bar l\rangle + |C\,b^2 l \bar a\rangle + 
|C\,b^2 a \bar l\rangle \over 2} \\
|\psi_2&(S\!g)\rangle =  
- 2\; {F_{al}^{(al)}F_{al}^{(l)} \over \Delta_1 U} \; {|C\,l^2 a^2\rangle + |C\,l^2 b^2\rangle \over \sqrt{2}} \\&
\qquad \quad \; - 2\sqrt{2} \; {F_{al}^{(al)}F_{al}^{(b)} \over \Delta_1 \Delta_2}\; |C\,a^2 b^2\rangle 
\end{align*}
The quality of the description of the effective exchange ${\cal
J}_{ab}$ will therefore be conditioned by the numerical accuracy of
the above effective first and then second order wave-function
coefficients. It is well known that renormalization of these
coefficients due to the dynamical polarization and correlation effects
are dominated by the screening effects; this is the dynamical
repolarization of the different above
configurations~\cite{poldyn}. One can therefore expect that the
dynamical repolarization of ligand-to-metal charge-transfer
configurations will be crucial for the accuracy of the effective
exchange.

\section{On the dynamical repolarization of the ligand-to-metal charge-transfer configurations} 
\label{sec3}

The Difference Dedicated Configurations Interaction (DDCI) method have
been very successful in accurately predicting the effective magnetic
exchange integrals. In order to propose a computationally less
demanding but equally accurate method, it is of great interest to
analyze how the above cited ligand-to-metal charge-transfer and how their
corresponding dynamical polarization contributions are accounted for
in the DDCI method.

\begin{figure}[h]
\resizebox{8cm}{4cm}{\includegraphics{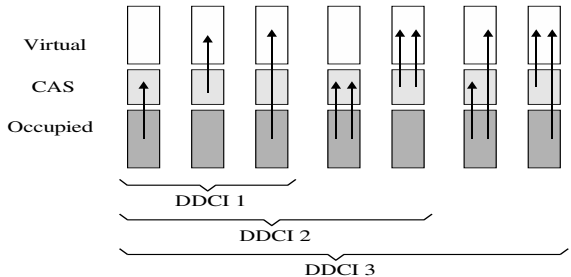}}
\vspace*{0.5eM}
\caption{Schematic representation of the DDCI1, DDCI2 and DDCI3 or DDCI
configurations.}
\label{f:DDCI}
\end{figure}

Let us now briefly recall the main principles of the DDCI or DDCI3 and
of the DDCI1 and DDCI2 simplified versions. The DDCI starts from a
multi-reference configuration interaction (CI) wave functions
containing all single and double excitations that can be constructed
from all determinants defining a minimal complete active space
(CAS). This minimal CAS does only contains the {\em magnetically
active orbitals and electrons} and therefore while it includes the
superexchange mechanisms it does not account for the through bridge
interactions.  The DDCI method is based on the restriction of the CI
space to all single and double excitations on the CAS determinants
which do contribute to the effective exchange integral (or
equivalently singlet-triplet excitation energy) at the second order of
perturbation (see figure~\ref{f:DDCI}).
The DDCI (often referred to as DDCI3) space therefore include all
double-excitations on the CAS determinants ---~complemented by the
necessary determinants to have spin- and symmetry-adapted
configurations~--- except the 2-holes 2-particles determinants. Hence,
the double-excitations from two doubly-occupied orbitals toward two
virtual ones are not included in the DDCI space. The DDCI2 space
excludes in addition all 2-holes 1-particle and 1-particle 2-holes
excitations (see figure~\ref{f:DDCI}). Finally the DDCI1 space is the
configuration space of all single excitations on the CAS ---~of course
again completed for spin and space symmetry.  One immediately sees
that the reference determinants involved in $|\Psi_0(S\!g)\rangle$ and
$|\Psi_0(T\!p)\rangle$ are included in the CAS, as well as the metal
ionic configurations. The ligand-to-metal charge-transfer
configurations already appear in the DDCI1 space, while the double
ligand-to-metal charge-transfers appear in the DDCI2 space. The
dynamical polarization effects on these configurations correspond
precisely to the single excitations. Therefore these effects enter at
the DDCI1 level for the reference determinants as well as for the
metal ionic determinants. The dynamical polarization of the
ligand-to-metal charge-transfer configurations are spread over the
DDCI1, DDCI2 and DDCI3 (the most numerous) spaces (see appendix
B). Finally, the dynamical polarization of the double ligand-to-metal
charge-transfer configurations are not taken into account in the DDCI
space. The very good results of the DDCI method, despite the lack of
the double transfers screening effects, can be attributed to the fact
that these double transfers have usually a negligible contribution to
the effective exchange process (their weight is for instance smaller
than $2.1\;10^{-2}$ in the triplet or singlet wave function in the
case of the copper oxide compounds).

Table~\ref{t:ddci23} details the effective exchange integral on a few
copper oxide and nickel fluoride systems as computed using the DDCI2
and DDCI3 methods (see appendix A for computational details). The
difference between the DDCI2 and DDCI3 results range between $30\%$ to
$40\%$ of the nominal value of the exchange integral on the copper
oxides and around $20\%$ on the nickel fluorides. These large differences
are in agreement with the above analysis on the crucial role of the
ligand-to-metal charge-transfers repolarization which largest
contributions are treated at the DDCI3 level only (see
appendix B).
\begin{table}[h]
\begin{tabular}{l|dddd}
Compound & \multicolumn{1}{c}{DDCI1} &  \multicolumn{1}{c}{DDCI2} &  \multicolumn{1}{c}{DDCI3} &   \multicolumn{1}{c}{Exp.} \\ 
\hline
$L\!a_2C\!uO_4$ & -91.1 & -96.3 & -145.5 & -135. \pm 5~\footnote{see ref.~\cite{JexpLa}} \\
$H\!g B\!a_2 C\!u O_4$ & -87.3 & -92.5 & -153.7 & \\
$S\!r_2C\!uO_2C\!l_2$  & -69.3 & -73.3 &  -131.0 & -125~\footnote{see ref.~\cite{JexpSr}} \\ 
$T\!l B\!a_2 C\!u O_5$  & -95.6 & -101.4 & -166.7 & \\
\hline
$K N\!i F_3$ & -5.09 & -5.38 & -6.82 & -8.2~\footnote{see ref.~\cite{JexpK}} \\
&  -5.03  & -5.30  &  -6.64 & \\[1ex]
$K_2 N\!i F_4$ & -5.52 & -5.84 & -7.34 & -8.6~\footnote{see ref.~\cite{JexpK2}} \\
&  -5.46  & -5.76  &  -7.18 & \\ 

\end{tabular}
\caption{DDCI2 versus DDCI3 evaluation of the effective exchange
integrals between copper or nickel atoms (in meV). In the copper
compounds the exchange integral as been computed as the singlet-triplet
local excitation energy, while for the nickel compounds, due to the
spin 1 character of the nickel atoms, the exchange integral has been
computed both from the triplet-quintet excitation energy (first line)
and from the singlet-triplet excitation energy (second line). It can 
be noted that both methods yield equivalent results.}
\label{t:ddci23}
\end{table}
 
One can therefore suggest to include the ligand-to-metal charge-transfer 
configurations already in the CAS. This is simply achieved by adding the
bridging ligand orbital to the CAS. In this way the repolarization of the
ligand-to-metal charge transfers configurations as well as the
repolarization of the ligand-to-metal double charge transfers would be
accounted for at the DDCI1 level. 

Looking at the CI space sizes, the size of the DDCI1 space scales as
$n_{CAS}\times n_{occ}\times n_{virt}$, while the DDCI space scales as
$n_{CAS} \times n_{orb} \times n_{occ} \times n_{virt}$ where
$n_{CAS}$, $n_{orb}$ and $n_{occ}$ have the same meaning as in the
introduction section and $n_{virt}$ is the number of virtual orbitals. It
immediately follows that in addition to a better treatment of the
ligand-to-metal charge transfer excitations, the DDCI1 method built up
on an extended CAS will be less computationally demanding despite the
increase of the CAS size.

\section{Results}
\label{sec4}

The key question is now how to determine a systematic procedure to
select, within the doubly occupied orbitals, the pertinent ligand
orbital(s) that mediate the effective exchange interactions. This is
not an easy task because, although the line of reasoning in the
previous section follows a localized orbital scheme, the actual
calculations are carried out in a delocalized basis as usual in ab
initio calculations. 

In the DDCI procedure the orbitals are usually defined using a minimal
CASSCF procedure for the high spin state. We will stick on this point
and extract from the minimal CAS inactive orbitals the pertinent
ligand ones. The {\em magnetically active orbitals} are essentially of
atomic nature ($nd$ orbitals of the metal atoms) and thus the ligand
orbitals optimally mediating the exchange interaction should i) exhibit a
large overlap with these $nd$ atomic orbitals (this ensures a
large numerator in eq.~\ref{eq:J}), ii) be close enough to the Fermi
level in order to preserve the perturbative expression (small
denominators also in eq.~\ref{eq:J}). 

Table~\ref{t:rec}  shows the overlap between the copper $3d_{x^2-y^2}$
(resp. nickel $3d_{z^2}$ and $3d_{x^2-y^2}$) atomic
orbitals and the highest energy doubly occupied orbitals,
those located immediately below the Fermi level. 

\begin{table}[h] 
\hspace*{-1eM} 
\begin{tabular}{l|c|d@{\hspace{0.3ex}}d@{\hspace{0.3ex}}d@{\hspace{0.3ex}}d@{\hspace{0.3ex}}d|d@{\hspace{0.3ex}}d@{\hspace{0.3ex}}d@{\hspace{0.3ex}}d@{\hspace{0.3ex}}d@{\hspace{0.3ex}}d} 
Compound & AO & \multicolumn{11}{c}{Overlap square $\times 10^{3}$} \\
         &    & \multicolumn{5}{c|}{$a_g$ irrep.} & \multicolumn{5}{c}{$b_{2u}$ irrep.} \\
\hline
$L\!a_2C\!uO_4$ &  $d_{x^2-y^2}$       & 1 &\bf 18 & 3 & & & 2 & 1 &\bf 31 & 1 & & \\ \hline
$H\!g B\!a_2 C\!u O_4$ & $d_{x^2-y^2}$ & 2 &\bf 28 & 3 & & & 1 & 2 &\bf 46 & 1 & & \\ \hline
$S\!r_2C\!uO_2C\!l_2$  & $d_{x^2-y^2}$ & 2 &\bf 23 & 4 & & & 2 & 2 &\bf 43 & 1 & & \\ \hline
$T\!l B\!a_2 C\!u O_5$ & $d_{x^2-y^2}$ & 5 &\bf 29 & 4 & & & 3 & 19&\bf 41 & 0 & & \\ \hline
$K N\!i F_3$ & $d_{x^2-y^2}$  & 0 &\bf 4 &\bf  3 & 1 & 0 & 1 & 0 &\bf 11 &\bf  3 & 0 & 0 \\
             & $d_{z^2}$      & 0 &\bf 1 &\bf  8 & 0 & 0 & 0 & 0 &\bf  4 &\bf  9 & 0 & 0 \\ \hline
$K_2 N\!i F_4$& $d_{x^2-y^2}$ & 1 &\bf 6 &\bf  1 & 0 & 0 & 1 & 1 &\bf 13 &\bf  0 & 0 & 0 \\
              & $d_{z^2}$     & 0 &\bf 0 &\bf 13 & 1 & 0 & 1 & 0 &\bf  0 &\bf 17 & 0 & 0 \\
\end{tabular}
\caption{ $10^{3}$ the square of the overlap between the atomic $3d$ orbital 
of the {\em magnetic atoms} as specified in the second column and the
first doubly-occupied orbitals below the Fermi level. Only orbitals
belonging to the same irreducible representation as the {\em
magnetically active orbitals}, namely $a_g$ and $b_{2u}$, have been
considered. The orbitals are ordered in increasing energetic order
from left to right within each irreducible representation.}
\label{t:rec}
\end{table}

\begin{figure}[h]
\resizebox{6cm}{8cm}{\includegraphics{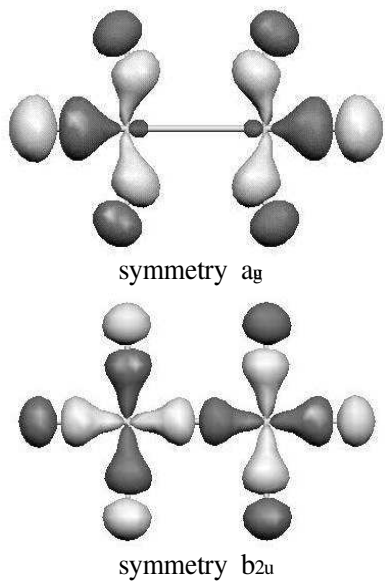}}
\vspace*{0.5eM}
\caption{Ligand orbitals of the copper oxide compounds having a
non-negligible overlap with the 3d atomic orbitals of the
copper. These orbitals are composed of atomic 2p orbitals of the
oxygens, pointing toward the cooper atoms }
\label{f:lorb}
\end{figure}

One immediately notices that in most cases, except for the $T\!l
B\!a_2 C\!u O_5$ compound, there is one ligand orbital per symmetry
that exhibits a non-negligible overlap with each of the considered
$3d$ atomic orbitals. The choice of the bridging ligand orbital to add
to the CAS appears thus to be quite unambiguous (two orbitals for the
copper compounds and four for the nickel ones).  At this point one
should notice that, unlike the usual description of the mediated
effective exchange, there is not only one bridging orbital that
contribute (two for $S=1$ {\em magnetic atoms}) but one per 
{\em magnetic orbital}.
For the case of the $La_2CuO_4$ copper oxide compound, these orbitals
are represented in figure~\ref{f:lorb}.  While the $b_{2u}$ orbital
has a large coefficient of the bridging oxygen $p_x$ orbital, this is
not the case for the $a_g$ orbital that does not have any significant
weight on any orbital of the bridging oxygen.  Yet, this $a_g$
orbital has to be considered, since, according to our results, the
antiferromagnetism is otherwise strongly overestimated. Indeed, a
DDCI2 calculation on the minimal CAS yields an effective exchange
integral of $-86 meV$.  The DDCI2 result improves to $-141meV$ when
both ligand $a_g$ and $b_{2u}$ orbitals are included in the active
space, but deteriorates to $-196 meV$ when solely the ligand $b_{2u}$
orbital is added and to $-39 meV$ when solely the ligand $a_{g}$
orbital is added.

Consequently, DDCI1 calculations have been performed using CAS spaces
enlarged to the two ligand orbitals for the copper oxides and the four
ligand orbitals for the nickel fluoride (see appendix A for
computational details). Results are reported in table~\ref{t:res} and
compared with the DDCI3 reference calculations using the minimal CAS.
\begin{table}[h] 
\hspace*{-1eM} 
\begin{tabular}{l|d|d|d} 
Compound  & \multicolumn{1}{c|}{minimal CAS} & \multicolumn{1}{c|}{extended CAS} & \multicolumn{1}{c}{exp.} \\
 & \multicolumn{1}{c|}{DDCI 3} & \multicolumn{1}{c|}{DDCI 1} &\\
\hline
$L\!a_2C\!uO_4$        & -145.5 & -148.9 & -135.5 \pm 5\\
$H\!g B\!a_2 C\!u O_4$ & -153.7 & -155.8 & \\
$S\!r_2C\!uO_2C\!l_2$  & -131.0 & -138.7 & -125 \\
$T\!l B\!a_2 C\!u O_5$ & -166.7 & -125.2 & \\ \hline
$K N\!i F_3$   & -6.82 & -7.20  & -8.2 \\
               & -6.64 & -7.08  & \\[1ex]
$K_2 N\!i F_4$ & -7.34 & -7.62 & -8.6 \\
               & -7.18 & -7.51 & \\
\end{tabular}
\caption{Extended CAS DDCI1  in meV and minimal CAS DDCI3 evaluations of the
effective exchange integrals. For the nickel compound the first line
corresponds to the effective exchange evaluated from the
triplet-quintet excitation energy, while the second line corresponds
to the evaluation from the singlet-triplet excitation energy. }
\label{t:res}
\end{table}
From the results in table~\ref{t:res}, it is clear that DDCI3 
using minimal CAS as reference for single and double excitations and the DDCI1
based on the extended CAS reference essentially yield the same results,
except for the $T\!l B\!a_2 C\!u O_5$ compound. 
Indeed, in the other copper oxides, the
relative error between the two methods is always smaller than $6\%$, 
being as weak as $1.3\%$ for the $H\!g B\!a_2 C\!u O_4$ compound.  These
results therefore confirm the previous analysis that the crucial physical
effect to be accounted for in the DDCI3 wave function is the dynamical
repolarization of the ligand-to-metal charge transfer
configurations. Nevertheless, it is noticeable that the only case giving lesser
quality results is the case for which it was not possible to clearly
identify the bridging ligand orbitals. 

In the nickel compounds the relative error between the two methods is
somewhat larger due to the weakness of the exchange. The absolute
error is however only of about half a meV with the DDCI1 results being
slightly closer to the experimental values. The improvement can be
understood by the fact that the repolarization of the double
ligand-to-metal charge transfer is accounted for in the
extended-CAS-DDCI1 calculation while it is not in the
minimal-CAS-DDCI3 one. It is worth pointing out that in the nickel
compounds the fact that four {\em magnetic orbitals} and four ligand
orbitals are involved increases considerably the number and importance
of double ligand-to-metal charge transfer.

\section{Conclusion}
\label{sec5}
From a careful analysis of the mechanism of the through-bridge
effective exchange interactions between two magnetic atoms, we have
shown that one of the crucial effects in the effective exchange
mechanism is the ligand to metal charge transfer. From this analysis,
it appears that an accurate evaluation of the effective exchange
integral (or equivalently of the singlet-triplet excitation energy)
requires a proper treatment of the dynamical repolarization on these
charge transfer configurations. To this end a new method has been
proposed for the determination of the bridging ligand orbitals 
involved in these charge transfer excitations.  It is therefore
proposed to include these mechanisms already in the zeroth-order wave
function. To this end it is enough to include the bridging ligand
orbitals in the complete active space (CAS). It is noticeable, that on
a rather large series of compounds the evaluation of the effective
exchange using the traditional minimal-CAS-DDCI3 method and using the
present enlarged-CAS-DDCI1 approach yield almost identical results
provided that the bridging ligand orbitals can be determined without
ambiguity. 

There is a very important practical consequence of the present
study. The final dimension of the configuration interaction space in
the enlarged-CAS-DDCI1 strategy is considerably smaller than the one
corresponding to the traditional minimal-CAS-DDCI3. This permits to
lift the bottleneck encountered in the DDCI method when applied to
systems with more magnetic centers or more magnetic orbital and
electrons per center, i.e. the very large dimension of the CI space
leading to almost intractable diagonalization problems. Indeed, 
in most systems, the number of single excitations out of the
extended CAS are much less numerous than the DDCI space on the minimal
CAS. In cases where the double ligand-to-metal charge transfers are
non negligible, one can even expect the enlarged-CAS-DDCI1 method to even 
yield better results than the minimal-CAS-DDCI3. Indeed, the
repolarization of these double transfer excitations is properly taken
into account in the former method while they are totally omitted in
the latter.

To conclude~: the present study points out the crucial importance of the
repolarization of the charge transfer and double charge transfer
excitations, in order to accurately determine the effective exchange
(or magnetic coupling) constants, and it proposes a simple yet efficient
method to include this effects. It is suggested that this new procedure
will permit the study of systems with a larger number of magnetic
electrons and orbitals such as the technologically relevant $LaMnO_3$
related compounds.

\section{acknowledgments}
 {\sf This research has been supported by the Spanish DGICYT grant
BQU2002-04029-CO2-01 and, in part, by Generalitat de Catalunya grant
2001SGR-00043. 
The authors are indebted to Dr. D. Maynau for providing
us with the last version of the CASDI set of programs used in most of
the presented calculations. 
Computer time was provided by the Centre
de Supercomputaci\'o de Catalunya, CESCA, and Centre Europeu de
Paral.lelisme de Barcelona, CEPBA, through IHP, program under contract
HPRI-CT-1999-00071 held by the CESCA/CEBPA. 
A.G. is grateful to the
European Community and the Universit\'e Paul Sabatier for financing
his stay in Barcelona through the above mentioned IHP program and
ATUPS programs.
M.L.M. gratefully acknowledges grant LN00A016 from the Ministry of
Education of the Czech Republic and a scholarship from the French
government that enabled her stay in Toulouse.}

\appendix

\section{Appendix A}

\label{a:a}

The $L\!a_2C\!uO_4$, $H\!g B\!a_2 C\!u O_4$, $T\!l B\!a_2 C\!u O_5$,
$S\!r_2C\!uO_2C\!l_2$ compounds are of perovskite geometry. The copper
atoms are localized in $C\!uO_2$ planes, responsible for the
conduction and supra-conduction properties of the systems. These
compounds are essentially ionic crystals, the copper being in the
$C\!u^{2+}$ oxidation state with a $3d^94s^0$ configuration. The ligand field
splitting results in a magnetic non-degenerated $3d_{x^2-y^2}$
orbital.

The nickel-based compounds exhibit similar perovskite structure.  The
nickel atoms exhibit a $N\!i^{2+}$ oxidation state with a high-spin ($S=1$)
$3d^8$ atomic configuration. In this case the magnetic orbitals are
the two $e_g$ orbitals of the $O_h$ symmetry group. 

The geometry of the systems computed here have been extracted from X-ray
experimental data~\cite{geom}. The embedding have been derived in
order to account for the major effects of the rest of the crystal on
the cluster under consideration, namely the electrostatic potential,
the exclusion effects of the electrons of the rest of the crystal on
the computed fragment orbitals. 

The metal atoms are treated using core pseudo-potentials and valence
$3-\zeta$ quality basis set~\cite{CuNi}. The bridging atoms are
treated using large ANO basis set~\cite{OF} and the other ligands
using effective core potentials and valence $3-\zeta$ quality basis
set~\cite{OF2}. 

The CASSCF calculations have been performed using the MOLCAS
Version 5.2 package~\cite{molcas}.

\section{Appendix B}
\label{a:b}

\begin{figure}[h]
\resizebox{8.1cm}{11.5cm}{\includegraphics{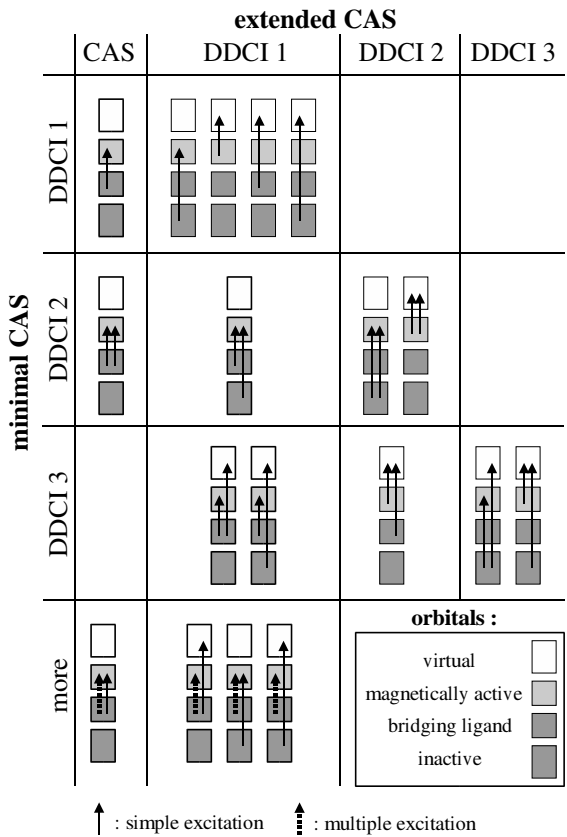}}
\vspace*{0.5eM}
\caption{Decomposition of the enlarged CAS DDCI1, DDCI2 and DDCI3
spaces on their minimal CAS counterparts.}
\label{f:gdpt}
\end{figure}

Figure~\ref{f:gdpt} displays the different configurations involved in
the enlarged CAS DDCI1, DDCI2 and DDCI3 spaces as a function of the
minimal CAS DDCI1, DDCI2 and DDCI3 spaces.


\begin{thebibliography}{9999}


\bibitem{htc} See for instance~:
J. G. Bednorz and K. A. M\"uller, {\it Z. Phys.} {\bf B 64}, 189 (1986)~;
J.G. Bednorz, {\em Earlier and Recent Aspects of
Superconductivity}, Eds. K.A. M\"uller, Springer, Berlin (1990)~;
E. Dagoto in~: {\em Recent Progress in Many Body Theories},
vol. {\bf 4}, Eds. E. Schachinger, H. Mitter and M. Sormann, Plenum,
New-York (1995).

\bibitem{mang} For reviews see for instance~: \\
V. Kiryukhin, D. Casa, J.P. Hill, B. Kelmer, A. Vigliante, Y. Tomioka
and Y. Tokura, {\it Nature} {\bf 386}, 813 (1997)~; A. Asamitsu, Y. Tomioka,
H. Kuwahara and Y. Tokura, {\it Nature} {\bf 388}, 50 (1997)~; M. F\"ath,
S. Freisem, A.A. Menovsky, Y. Tomioka, J. Aarts and J.A. Mydosh,
{\it Science} {\bf 285}, 1540 (1999)~;
T. Kimura and Y. Tokura, {\it Annu. rev. mater. sci.} {\bf 30}, 451 (2000)~;
Y.K. Yoo, F. Duewer, J.W. Haltao Yang, J.W. Dong Yi, J.W. Li and
X.D. Xiang, {\it Nature} {\bf 406}, 704 (2000).


\bibitem{phmgmol} 
V. Marvaud, J.M. Herrera, T. Barilero, F. Tuyeras, R. Garde,
A. Scuiller, C. Decroix, M. Cantuel, C. Desplanches, invited review,
{\it Monatshefte für Chemie} {\bf 134}, 149 (2003). 

\bibitem{phmg} O. Sato, T. Iyoda, A. Fujishima and K. Hashimoto, {\it
Science} {\bf 271}, 49 (1996)~; M. Verdaguer, {\it Science} {\bf 272},
698 (1996)~; O. Sato, T. Iyoda, A. Fujishima and K. Hashimoto, {\it
Science} {\bf 272}, 704 (1996).


\bibitem{book-olivier-kahn} O. Kahn, {\it Molecular magnetism},
Wiley-VCH (1993). 


\bibitem{dft} See for instance~: 
W.E. Pickett,  {\it Rev. Mod. Phys.} {\bf 61}, 433 (1989)~; 
I. de P.R. Moreira and F. Illas, {\it  Phys. Rev.} {\bf B 60}, 5179 (1999). 


\bibitem{revue} M.-B. Lepetit, {\it Recent Research Develepments in
Quantum Chemistry}, p. 143, Vol. 3, Transworld research Network
(2002).


\bibitem{HEFF} I. de P.R. Moreira, N. Suaud, N. Guihery, J.P. Malrieu,
R. Caballol, J.M. Bofill and F. Illas, {\it Phys. Rev.} {\bf B 66}, 134430 
(2002).  


\bibitem{heis}
W. Heisenberg, {\it Z. Phys.}, {\bf 49}, 619 (1928)~;
P. A. M. Dirac, {\it Proc. R. Soc. London}, {\bf A 123}, 714 (1929)~;
J. H. Van Vleck, {\it The Theory of Electric and Magnetic
Susceptibilities.}  Oxford University Press, Oxford (1932).

\bibitem{t-J} C. Gros, R. Joynt and T. M. Rice, {\it Phys. Rev.}, {\bf B 36},
381 (1987)~; P. W. Anderson, {\bf Science}, {\bf 235}, 1196 (1987)~;
F. C. Zhang and T. M. Rice, {\it Phys. Rev.}, {\bf B 37}, 3759 (1988).

\bibitem{CALZADO99} I. de P.R. Moreira, F. Illas, C.J. Calzado,
J.F. Sanz, J.P. Malrieu, N. Ben Amor and D. Maynau, {\it Phys. Rev.}
{\bf B 59}, R659 (1999). 

\bibitem{NiO} I. de P.R. Moreira, F. Illas and R.L. Martin, {\it
Phys. Rev.} {\bf B 65}, 155102 (2002).

\bibitem{dft1} R.L Martin et F. Illas, Phys. Rev. Letters {\bf 79}, 1539
(1997). 

\bibitem{la2cuo4} B. Keimer, N. Belk, R.J. Birgeneau, A. Cassanho, 
C.Y. Chen, M. Greven, M.A. Kastner, A. Aharony, Y. Endoh, R.W. Erwin
and G. Shirane, Phys. Rev. {\bf B 46} 14034 (1992).

\bibitem{ddci} J. Miralles, J. P. Daudey and
R. Caballol, {\it Chem. Phys. Lett.} {\bf 198}, 555 (1992)~;
V. M. Garc\'ia, O. Castell, R. Caballol and J. P. Malrieu, {\it
Chem. Phys. Lett.} {\bf 238}, 222 (1995); V. M. Garc\'ia, M. Reguero
and R. Caballol, {\it Theor. Chem. Acc.} {\bf 98}, 50 (1997).

\bibitem{molmagn} J. Cabrero {\it et al.}, J. Phys. Chem. {\bf A 104}, 
9983 (2000). 

\bibitem{xino}
D. Mu\~noz, F. Illas and I. de P.R. Moreira, Phys. Rev. Letters {\bf
84}, 1579 (2000).


\bibitem{vana1} N. Suaud and M.-B. Lepetit, Phys. Rev. {\bf B 62} 402 (2000). 

\bibitem{xinoloc}  F. Illas, I. de P.R. Moreira, C. de Graaf, O. Castell
and J. Casanovas, Phys. Rev. {\bf B56}, 5069 (1997)~;
I. de P.R. Moreira, F. Illas, C. Calzado, J.F. Sanz J.-P. Malrieu,
N. Ben Amor and D. Maynau, Phys. Rev. {\bf B59}, R6593 (1999).

\bibitem{xinotca} F. Illas, I de P.R. Moreira, C. de Graaf and V. Barone, 
Theoret. Chem. Acc. {\bf 104}, 265 (2000).

\bibitem{qdpt} See for instance~: I. Lindgren and J. Morrison, {\em Atomic 
Many-body Theory}, Springer, Berlin (1982), and references therein.

\bibitem{en} P.S. Epstein, Phys. Rev. {\bf 28}, 695 (1926)~; R.K. Nesbet, 
Proc. R. Soc. London Ser. {\bf A 230}, 312 (1955)~; P. Claverie,
S. Diner and J.-P. Malrieu, Int. J. Quantum. Chem. {\bf 1}, 751 (1967).

\bibitem{poldyn} Ph. Hiberty, Chem. Phys. Letters {\bf 189}, 259 (1992)~; 
R. Ghailane, M.-B. Lepetit and J.-P. Malrieu, J. Phys. Chem. {\bf 97},
94 (1993).

\bibitem{JexpLa} B. Kreimer, N. Belk, R. J. Birgeneau, A. Cassanho,
C. Y. Chen, M. Greven, M. A. Kastner, A. Aharony, Y. Endoh,
R. W. Rewin and G. Shirane, Phys. Rev. {\bf B 46}, 14 034 (1992).

\bibitem{JexpSr} D. Vaknin, S. K. Sinha, C. Stassis, L. L. Miller2 and
D. C. Johnston, Phys. Rev. {\bf B 41}, 1 926 (1990).

\bibitem{JexpK}  M. E. Lines, Phys. rev. {\bf 164}, 736 (1967).

\bibitem{JexpK2} L. J. de Jongh and R. Miedema, Adv. Phys. {\bf 23}, 1 (1974).


\bibitem{geom} $L\!a_2C\!uO_4$~: J. M. Longo and P. M. Raccah,
J. Solid State Chem. {\bf B 6}, 526 (1973). \\
$H\!g B\!a_2 C\!uO_4$~: S. N. Putilin, E. V. Antipov, O. Chmaissen and
M Marezio, Nature (London) {\bf 362}, 226 (1993). \\
$T\!l B\!a_2 C\!u O_5$~: L. F. Mattheiss, Phys. Rev. {\bf B 42} 10 108
(1990). \\
$S\!r_2C\!uO_2C\!l_2$~: L. L. Miller, X. L. Wang, C. Stassis,
D.C. Johnston, J. Faber, and C.-K. Loong, Phys. Rev. {\bf B 41}, 1921
(1990).


\bibitem{CuNi} P.J. Hay and W.R. Wadt, J. Chem. Phys. 82 (1985) 299.

\bibitem{OF} K. Pierloot, B. Dumez, P.-O. Widmark and B. O. Roos
Manuscript in preparation. 

\bibitem{OF2} Ph. Durand and J.C. Barthelat, Theor. Chim. Acta {\bf
38}, 283 (1975).  


\bibitem{molcas} ${\cal MOLCAS}$ Version 5.2,
K. Andersson, 
M. Barysz,
A. Bernhardsson,
M. R .A. Blomberg, 
Y. Carissan,
D. L. Cooper, 
M. Cossi,
T. Fleig,
M. P. F\"ulscher, 
L. Gagliardi,
C. de Graaf,
B. A. Hess,
G. Karlstr\"om,
R. Lindh, 
P.-{\AA}. Malmqvist, 
P. Neogr\'{a}dy, 
J. Olsen, 
B. O. Roos, 
B. Schimmelpfennig,
M. Sch\"utz, 
L. Seijo, 
L. Serrano-Andr\'es, 
P. E. M. Siegbahn, 
J. St{\aa}lring,
T. Thorsteinsson,
V. Veryazov,
M. Wierzbowska,
and P.-O. Widmark, Lund University, Sweden (2001).


\end{thebibliography}
\end{document}